\documentstyle[epsf,prd,aps,floats]{revtex} 
\input{psfig}
  
\newcommand{\be}{\begin{equation}}   
\newcommand{\ee}{\end{equation}}   
\newcommand{\bea}{\begin{eqnarray}}   
\newcommand{\eea}{\end{eqnarray}}

\begin{document}    
\twocolumn[\hsize\textwidth\columnwidth\hsize\csname@twocolumnfalse\endcsname    
\author{Stephon Alexander and Jo\~ao Magueijo}   
\date{\today} 
\address{Theoretical Physics, The Blackett Laboratory, 
Imperial College, Prince Consort Road, London, SW7 2BZ, U.K.} 
\title{Non-commutative geometry as a realization of
varying speed of light cosmology}   
\maketitle    
\begin{abstract} 
We examine the cosmological implications of space-time non-commutativity,
discovering yet another realization of the varying speed of light model.
Our starting point is the well-known fact that non-commutativity
leads to deformed dispersion relations, relating energy and momentum,
implying a frequency dependent speed of light. A Hot Big Bang Universe
therefore experiences a higher speed of light as it gets  hotter.
We study the statistical physics of this ``deformed radiation'', recovering
standard results at low temperatures, but a number of novelties at high
temperatures: a deformed Planck's spectrum, a temperature dependent
equation of state $w=p/\rho$ (ranging from $1/3$ to infinity), 
a new Stephan-Boltzmann
law, and a new entropy relation.  These new photon properties closely
mimic those of phonons in crystals, hardly a surprising analogy.
They  combine to solve the horizon and flatness problems, explaining also the
large entropy of the Universe. We also show how
one would find a direct imprint of non-commutativity in the spectrum
of a cosmic graviton background, should it ever be detected. 
\end{abstract}   
\pacs{PACS Numbers: *** } 
 ]

\renewcommand{\thefootnote}{\arabic{footnote}} 
\setcounter{footnote}{0} 

\section{Statement of purpose}
Inflation and string theory are widely perceived as the leading
schools of thought in cosmology and fundamental physics, 
respectively. And yet finding common ground between them, 
say by deriving inflation from a concrete string theory model,
has so far remained elusive. Such
an enterprise is far from frivolous, since cosmology is probably
the most realistic laboratory for testing string theory. 
For this reason it is disappointing that after all 
these years, and following a number of radical revolutions in both fields,
the two theories have still failed to condense into a single construction.

Could it be that somehow we got things wrong on one side of 
the story? The inspirational origins of inflation are founded on a number 
of problems of Big Bang cosmology, namely the horizon, flatness,
homogeneity, and entropy problems \cite{infl1,infl2b,infl2,infl3}. 
A recently proposed alternative explanation
to these problems is the varying speed of light (VSL) 
theory~\cite{mof1,am,li}, the idea
that light travelled faster in the early Universe. Currently VSL
is far less developed than inflation, with full model building work
hardly started.  But could VSL be more
amenable to a direct connection with fundamental physics 
\cite{kir,alex}, more specifically with string theory?

In this paper we establish a connection between non-commutative
geometry and VSL. Over the last few years non-commutativity of space-time
coordinates has become part and parcel of any attempt to quantize 
gravity. Similar behaviour was already spelled out in Appendix I 
of \cite{am} (see also \cite{li}), where it was shown that under VSL
the partial derivatives $\partial_t$ and $\partial_x$ do not commute.
Furthermore non-commutative geometries are known to 
lead to deformed dispersion relations \cite{amel}, 
which imply a frequency-dependent speed of light. 

A deep connection is up for grabs, and such is the purpose of this paper.

\section{Cosmology and Non-Commutativity}\label{nonc}

One of the pillars of the standard Big Bang (SBB)
 is the theory of General Relativity. The
problems of the SBB model suggest that classical General Relativity
is inconsistent in
the very early universe where quantum corrections to gravity become
significant.  Interestingly, most attempts toward a theory of quantum
gravity such as string/M-theory realize that space-time itself is
non-commutative.  In particular, non-commutativity is concretely realized
in a non-perturbative formulation of M-theory, (M)atrix theory, where the D-0 
brane collective coordinates are in general
non-commuting; this is interpreted as a non-commuting space-time.  In the infinite momentum frame (IMF), non-commuting space-time
emerges from (M)atrix theory \cite{lenny}.  Also, in the open string sector of
string theory, non-commutative geometry arises when the NS-B field is
turned on; the open string end points become  ``polarized'' and hence
non-commuting \cite{SJ,SW}.  In the low energy limit of open string
theory in the presence of a B field, these dipoles are realized as
non-commutative solitons in  Non-commutative Yang-Mills
(NCYM) theory where they exhibit a VSL behavior \cite{Hash}.

The coordinate operators of a non-commuting space obey the following
commutation relation.
\be [\hat{x}^\mu,\hat{x}^{\nu}] = i \Theta^{\mu\nu} \label{thetaa} \ee
where $\Theta^{\mu\nu}$ is in general constant, antisymmetric, and  with
dimension $[length]^{2}$.  There exists a general prescription for
studying quantum field theories in non-commutative spaces with flat
backgrounds.  By simply taking a classical field theory defined on a
commuting space and replacing the product operation by a $*$ product,
one obtains a non-commutative field theory. The $*$ product is defined
from 
\be
\phi*\psi(x)=exp(\frac{i}{2}\Theta_{\mu\nu}\partial_{x_{\mu}}\partial_{x_{\nu}})\phi(x)\psi(y)|_{x=y}
\ee
\be = \phi(x)\psi(y) + \frac{i}{2}\Theta^{\mu\nu}\partial_{\mu}\phi
\partial_{\nu}\psi + \cal{O}\rm(\Theta^{2}) \ee
Notice that the $*$ product is non-commutative.  Also non-commutative
field theories are necessarily non-local since the $*$ product yields
an infinite number of derivatives in the Lagrangian.  This non-locality
in space-time, as we will see in the sections that follow, gives rise to a
VSL due to a modification in the dispersion relation of the quantum
fields in a non-commutative space-time.

Although non-commutativity is concretely realized in string theory,
space-space non-commutativity is better understood than its space-time
counterpart.  In both cases, however, non-commutativity is best
understood for constant B field (theta parameter); while more complicated versions of
non-commutativity in string theory generalized to non-constant B-field
and also its gravitational analogue
is currently under investigation\cite{ali,juan,SW}.  For VSL we are concerned
about space-time non-commutativity with non-constant B field ($\Theta$).

In light of the string-theoretic realization of non-commutativity, we
shall implement a more general algebra,  where the $\theta$ parameter is
not constant, but a linear function of $x$, a quantum deformed Poincare'
group. This extension is important because it precisely yields a VSL in
the photons by deforming their dispersion relations.  In equation (\ref{thetaa}) relativistic invariance remains classical and the
translations \be {\hat x}_{\mu} \rightarrow \hat{x}'_{\mu}=\hat{x}_{\mu} +
a_{\mu} \ee
preserving the algebra of (\ref{thetaa}) are commutative.  However, if
$\Theta_{\mu\nu}$ depends on $\hat{x}_{\mu}$,in particular if we assume
the linear $\hat{x}_{\mu}$ dependence, i.e.
\be
[\hat{x}_{\mu},\hat{x}_{\nu}]=ic^{\lambda}_{\mu\nu}\hat{x}_{\lambda}.\ee
then, the translations 
\be \hat{x}'_{\mu} \rightarrow \hat{x}_{\mu} + \hat{v}_{\mu} \ee
are not commutative.  For particular choices of the structure constants
$c^{\lambda}_{\mu\nu}$ this type of non-commutative structure falls
under the category of the quantum-deformed Poincare' group, with
non-commutative group parameters\footnote{For a more descriptive
classification of the quantum Poincare' groups see \cite{poin}}.  In
this work we consider a particular type of quantum Poincare' group; the
$\kappa$ deformation of space-time symmetries with non-commutativity of
space-time coordinates described by the following solvable Lie-algebra
relations:
\bea [\hat{x}_{0},\hat{x}_{i}]=\frac{i}{\kappa}\hat{x}_{i}, &\qquad
[\hat{x}_{l},\hat{x}_{j}]=0. \eea
Thus spatial translations still commute; non-commutativity affects 
only operations involving the time coordinate.

This type of non-commutativity has yet to be realized in string/M theory,
however we expect that it will be better understood in the future,
especially in the context of the non-commutative version of AdS/CFT
correspondence and in formulations where the B field ($\Theta$) 
is non-constant~\cite{Hash,juan}.

In the case of the deformed Poincar\'e algebra, the spatial
momenta remain commutative
\be [P_{i},P_{j}] = 0 \ee
and the rotation part of the Lorentz sector is left unchanged.  For
this reason we assume there is a scale $\Lambda_{NC}$
in the Early universe such that
\be \Theta= \frac{\kappa}{\Lambda^{2}_{NC}} \ee
where $\kappa$ is 
$\cal{O}\rm (1)$; ie. non-commutativity is strong.  As a result it is safe
to conjecture that since

(1) The $\kappa$ deformation does not act on the rotations and the
canonical momenta.

(2) The deformation is strong, $\kappa \sim \cal{O}\rm(1)$ ,

thus the non-commutativity can be described as a gas of massless
non-commutative radiation propagating in a conformally flat
Friedmann-Lemaitre-Robertson-Walker space-time.

The radiation is in
thermal equilibrium and is maximally correlated.  This is a statement
that the space-time is a Non-commutative FRW (NCFRW) with a gas of
photons.  The modified dispersion relation of the photons captures the
non-commutativity of the space-time.
The FRW space-time is endowed with flat spatial sections,
and is given by the line element  
\be ds^{2}= a^{2}(\eta)(-d\eta^{2} + dx^{2} + dy^{2} + dz^{2}). \ee
where $\eta$ is the conformal time and $(x,y,z)$ are spatial comoving
coordinates.  This metric is conformally flat and is locally equivalent
to the Minkowski metric $\eta_{\mu\nu}$.

Since our space-time is non-commutative, the notion of a continuous
space-time manifold is lost. In fact the separation between
energy-momentum and curvature, which is manifest in the Einstein field
equations is not as trivial. However, we can transcend this difficulty
by deforming the dispersion relations of the photons and study their
dynamics in a commuting space-time; this yields the nature of how matter
behaves in a quantized space-time.  So long as the photons are in
equilibrium, our approximation is valid.  

Therefore, our starting point for cosmology is a deformed Poincare group
whose action does not affect the conformal factor of the metric. A
similar approach was taken in the context of non-commutativity and
inflation \cite{brian}.
This is analogous to the standard hot
big bang scenario where our light degrees of freedom consist of
a gas of non-commutative radiation in thermal equilibrium.  
This radiation will exhibit a modified
dispersion relation which alters its equation of state.
 
\section{The thermodynamics of non-commutative radiation}

In the rest of the paper we will study the cosmological consequences of
modified dispersion relations for photons.  However, it is instructive
to provide a discussion of how modified dispersion relations arise in
the free field regime of the deformed $\kappa$ Poincair\'e group.  In
what follows, we will provide one of a few examples of a modified
frequency dependent dispersion relation.

In \cite{kappa} the authors were able to define at the perturbative
level a free field theory for massless bosons by identifying the
propagator and constructing the intertwiners between the representations
of the Poincare group acting on fields and states.  These intertwiners
are simply the wavefunctions of particles of definite spin. One can
define these intertwiners through the appropriate wave equations describing free
fields.

Therefore, the dispersion relations can be obtained from the invariant
wave operator on the $\kappa$-deformed Minkowski space
$\frac{\partial}{\partial\hat{x_{\mu}}}\frac{\partial}{\partial\hat{x_{\nu}}}$, which is
expressed in momentum space as 
\be C^{bcp}_{1}\left(1-\frac{C^{bcp}_{1}}{4\kappa^{2}}\right), \ee
where
\be C^{bcp}_{1}=(P^{2})e^{-P_{0}/\kappa} -
(2\kappa \sinh(\frac{P_{0}}{2\kappa})^{2} \ee
is the first Casimir of the
$\kappa$-deformed Poincaire algebra.

It follows that the spectrum of the modified massless wave operator
contains the deformed massless mode 
\be C^{bcp}_{1}\phi=0 \ee
which leads to the following dispersion relation
\be
k^{2}e^{-\omega/\kappa}-\left(2\kappa sinh(\frac{\omega}{2\kappa}\right)^{2}
=0
\ee
 The wave equation leading to the above dispersion relations is non-local
in time but can be reexpressed so that the dispersion relations
corresponds to an operator of second order in time derivatives and
non-local in space.
\be\label{dis} \omega^{2}=\left[\kappa log(1+\frac{k}{\kappa})\right]^{2} \ee.  

Let us now consider the following generalization of \ref{dis} for
 massless particles, for which:
\be\label{disp}
E^2 - p^2 c^2 f^2 =0
\ee
Here $f(E)$ gives rise to a frequency dependent 
speed of light, whereas $c$ is a possibly space-time dependent,
but frequency independent, speed. 
We thus factorize two different VSL effects previously considered
in the literature - a frequency dependent $c$
(as studied in \cite{amel}),  and a space-time dependent $c$ 
(as examined in \cite{mof1,am,li}). Most likely, the final
theory will contain both. Naturally the most general deformation 
need not factorize these two effects, or even produce a 
quadratic invariant as in Eqn.~(\ref{disp}),
but we shall retain this assumption. 

We shall consider two proposals for $f$, 
previously considered in the literature. Amelino-Camelia
and collaborators (eg. \cite{amel1}) have considered
\be\label{f1}
f=1+\lambda E
\ee
associated with the $\kappa$-Poincar\'e group, 
whereas following the quantum deformation of the Poincar\'e group
as studied by Majid one finds\cite{amel}\footnote{
We have applied $\lambda \rightarrow -2\lambda$ to the convention
used in \cite{amel} so that the 2 dispersion relations to be
considered agree to first order}:
\be
f={2\lambda E \over 1-e^{-2\lambda E}}
\ee
Here $\lambda$ is the deformation parameter and we have
set $\hbar=1$. The last deformation has also been proposed by
Kowalsky-Glikman. We will also consider a generalization of
(\ref{f1}), of the form
\be\label{f3}
f=(1+\lambda E)^\gamma
\ee
in particular for $2/3<\gamma<1$. The case $\gamma>1$
will be studied in detail elsewhere \cite{ncinfl}: it
leads to a realization of inflation, but not VSL.
We shall call these models 1, 2, and 3, respectively.

At once we find two possible definitions for the speed of light.
One may use the definition proposed in \cite{amel}:
\be
{\tilde c}= {dE\over dp}= {cf\over 1-{f'E\over f}}
\ee
(where ' denotes a derivative with respect to $E$)
for which model 1 gives:
\be
{\tilde c}=c(1+\lambda E)^2
\ee
whereas model 2 gives
\be
{\tilde c}= ce^{2\lambda E}\,.
\ee
Alternatively we may define the speed of light as
\be
{\hat c}={E\over p}= cf\,.
\ee
Both definitions, ${\tilde c}$ and ${\hat c}$,
play a role in what follows.

\subsection{The partition function and phase space densities}
We now proceed to 
examine the statistical physics of this ``deformed'' radiation.
We first note that the form of the partition function
\be
Z={\sum_r} e^{-\beta E_r}
\ee
does not depend on the dispersion relations; indeed it amounts to a 
definition of temperature based upon the fact that the thermal reservoir
has a much larger number of states than the system under examination.
Hence, for a boson gas,  the average number density 
of states with a given momentum still satisfies the Bose-Einstein
result: 
\be
n({\bf p})= {1\over e^{\beta E({\bf p})} -1}
\ee
since this formula depends only on the partition function and
the rules of counting associated with bosons. A similar undeformed
expression applies to fermions.

The density of momentum states for point particles is again unchanged,
as it simply reflects the use of periodic boundary conditions
(${\bf p}=(2\pi\hbar /L){\bf n}$ where ${\bf n}$ is a triplet of numbers,
and $L$ is the side of a given cubic volume).
All that changes is the relation between $E$ and $p$ and therefore 
the density of states $\Omega(E)$ with a given energy $E$  per
unit of volume. We find:
\be\label{densstat}
\Omega(E)={E^2\over \pi^2 \hbar ^3 {\hat c}^2 {\tilde c}}
={E^2\over \pi^2 (\hbar c)} {1\over f^3}{\left(
1-{f'E\over f}\right)}\,.
\ee
simply from $\Omega(E)dE=\Omega(p) dp$.

\subsection{The deformed Planck spectrum and the graviton background}
From $\rho(E)=n({\bf p}(E))\Omega(E)$ we thus obtain the
deformed Planck's spectrum:
\be\label{rohe}
\rho(E)={1\over \pi^2\hbar ^3 c^3}{E^3\over e^{\beta E}-1}
{1\over f^3}{\left(
1-{f'E\over f}\right)}\, .
\ee
One can easily show that only deformations of the form:
\be
f=[1+(\lambda E)^{-3}]^{-1/3}
\ee
leave the Planck spectrum unchanged. These are not very physical
in a cooling Universe, where the speed of light would initially 
be a constant, then start to drop, 
with $c\propto 1/a$, until nowadays.

All other dispersion relations affect Planck's law. This does not
conflict with experiment if $f\approx 1$ for $T\ll1/\lambda$. However
even then there may be an observational imprint. Gravitons decouple around
Planck time, and should constitute a thermal background similar to 
the cosmic microwave background (with a temperature of the same order). 
However their spectrum will be a deformed Planck spectrum, since it 
mimics the spectrum they had at the time they decoupled.
Hence if and when a graviton background is discovered its spectrum will
supply a direct measurement of the dispersion relations.

\begin{figure}
\centerline{\psfig{file=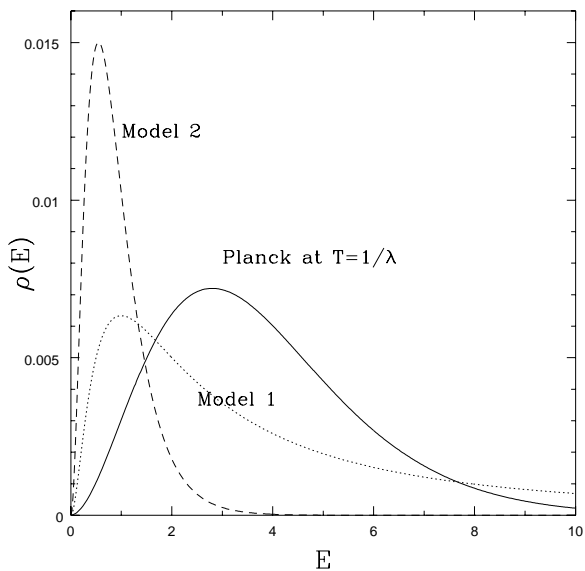,width=8 cm,angle=0}}
\caption{Deformed thermal spectra at high temperature ($T\lambda\gg 1$),
divided by $T^2$ (we have assumed $\lambda =1$). 
For reference we plotted a Planck spectrum at temperature
$T=1/\lambda$. For model 1 and 2 the color temperature saturates,
so that above $T=1/\lambda$ the peak in $\rho(E)$ does not
shift to higher energies.}
\label{fig1}
\end{figure}

More specifically model 1 leads to the deformed spectrum:
\be
\rho(E)={1\over \pi^2\hbar ^3 c^3}{E^3\over e^{\beta E}-1}
{1\over (1+\lambda E)^4}
\ee
For $T\ll1/\lambda$ distortions to this spectrum are negligible.
For high temperatures, though, the peak in $\rho(E)$
is not at $T$, but at $E=1/\lambda$ for all temperatures. In other
words the color temperature saturates at $T_{max}\approx 1/\lambda$. The
general form of the spectrum in this regime is
\be
\rho(E)={E^2 T\over \pi^2 (1+\lambda E)^4}
\ee
Model two leads to:
\be
\rho(E)={1\over \pi^2\hbar ^3 c^3}{E/\lambda^2\over e^{\beta E}-1}
e^{-4\lambda E}sinh^2(\lambda E)
\ee
which for $T\gg 1/\lambda$ leads to
\be
\rho(E)={T\over \lambda^2\pi^2}e^{-4\lambda E}sinh^2(\lambda E)
\ee
In both cases the spectrum is super-black (as opposed to gray); that
is, the overall amplitude of the spectrum with respect to black at
$T\approx 1/\lambda$ is enhanced by a factor of $T\lambda$.
In Fig.~\ref{fig1} we plot spectra for model 1 and 2, as well as
the undeformed Planck spectrum.

\begin{figure}
\centerline{\psfig{file=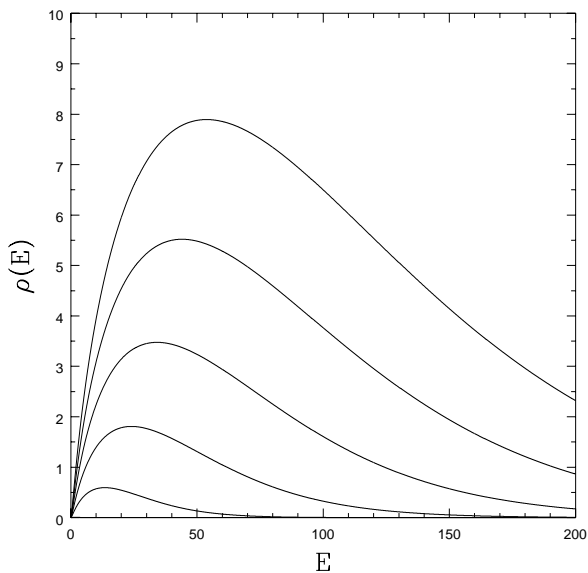,width=8 cm,angle=0}}
\caption{Deformed thermal spectra for model 3, with 
$\gamma=1/2$, as the temperature is increased. We see that for this
model at all temperatures, hotter radiation implies higher mean 
energies for photons, and a shift to the higher energies of
$\rho(E)$. }
\label{fig1.1}
\end{figure}

Model 3 is unique in that it never saturates the color temperature,
i.e. for all temperatures hotter radiation means a peak in the photons
distribution at higher energies. It's thermal spectrum is given by
\be\label{rhoe}
\rho(E)={1\over \pi^2\hbar ^3 c^3}{E^3\over e^{\beta E}-1}
{(1+(1-\gamma)(\lambda E)^\gamma)\over (1+(\lambda E)^\gamma)^4}
\ee
For $\gamma<2/3$ the peak of $\rho(E)$ scales like $T$. For
$2/3<\gamma<1$ the peak becomes very wide and covers energies
from $E\approx \lambda^{-1}$ to $E\approx T$.
We illustrate this feature  in Fig.~\ref{fig1.1}.

\subsection{The energy and entropy densities}
\begin{figure}
\centerline{\psfig{file=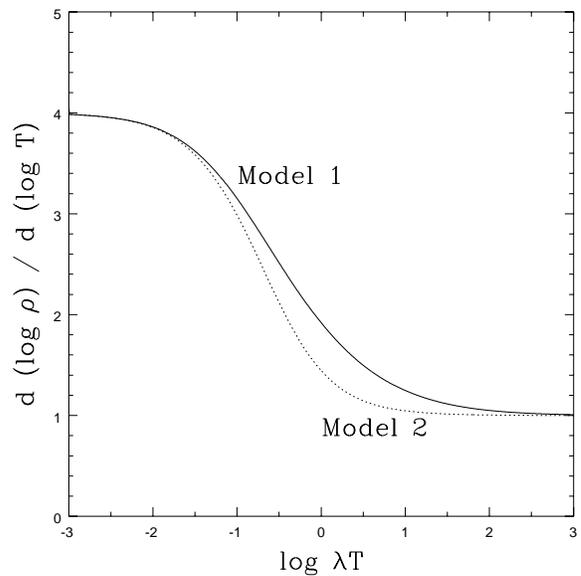,width=8 cm,angle=0}}
\caption{The transition from $\rho\propto T^4$ to $\rho\propto T$
at very high temperatures for non-commutative radiation.}
\label{fig2}
\end{figure}

Integrating $\rho(E)$ leads to a modified Stephan-Boltzmann
law. For low temperatures $\rho\propto T^4$ as usual, but
for $\lambda T\gg 1$ we find that $\rho\propto T$. This dependence
can be expressed in terms of the function 
\be
n(E)={d\log\rho\over d \log T}
\ee
which we plot in Fig.~\ref{fig2}. 
At low temperatures the proportionality constant is the usual
Stephan constant. At high temperatures the proportionality constant
is model dependent. In general, at high
temperature:
\be\label{stef}
\rho=\sigma^+ {T\over \lambda^3}
\ee
with
\bea
\sigma^+&=&{1\over \pi^2}\int_0^\infty  {x^2\over (1+x)^4}={1\over 3\pi^2}\\
\sigma^+&=& {1\over 8\pi^2}\int_0^1 (1-y^2)={1\over 24\pi^2}
\eea
for models 1 and 2, respectively. 
\begin{figure}
\centerline{\psfig{file=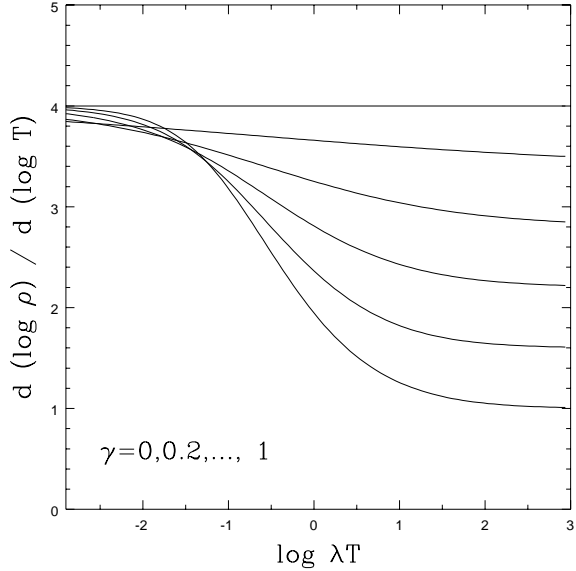,width=8 cm,angle=0}}
\caption{The transition from $\rho\propto T^4$ to $\rho\propto T^\zeta$
at very high temperatures for model 3. We see that for $\gamma<1$,
the high temperature relation has a power in the range $1<\zeta<4$.
In all cases hotter radiation translates into denser radiation. }
\label{fig2.1}
\end{figure}

For model 3 the situation is a bit different. The high energy 
Stephan-Boltzmann equation takes the form $\rho\propto T^\zeta$
with a power $\zeta$ in the range $1<\zeta<4$. An approximate
high temperature formula for $\zeta$ may be obtained by noting
that in Eqn.~\ref{rhoe} the last factor is approximately
$1/(\lambda T)^{3\gamma}$ over the most relevant parts of
the integrand. Hence we may expect at high energies
$\rho\propto T^{4-3\gamma}$ that is $\zeta\approx 4-3\gamma$.
We have verified numerically that this analytical approximation
works extremely well. In Fig.~\ref{fig2.1} we describe the transition
from the low to the high temperature behaviour for various
values of $\gamma$.

The entropy density then follows from 
\be
s=\int {dT\over T} {\partial \rho\over \partial T}
\ee
For low temperatures we recover the usual expression
\be\label{lowts}
s={4\over 3}\sigma T^3
\ee
whereas for high temperatures we now have
\be\label{hights}
s\approx \sigma^+\log{\lambda T}
\ee

\subsection{The equation of state}

The pressure formula (and so the equation of state)  
is also modified. As before it is given by a sum over all states
\be
p=\sum_s n_s {\left(-\partial E_s\over \partial V\right)}
\ee
but since now we have for each state
\be
E= {2\pi\hbar c f \over V^{1/3}}{\bf n}
\ee
we have 
\be
p={1\over 3 V}\sum_s n_s {E_s\over 1-{f'E\over f}}
={1\over 3}\int {\rho(E) dE\over 1-{f'E\over f}}
\neq {1\over 3}\rho
\ee
and so the equation of state of radiation is modified.

\begin{figure}
\centerline{\psfig{file=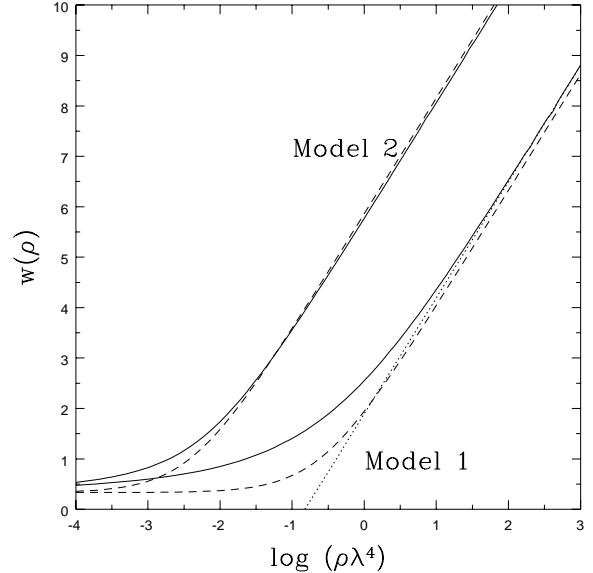,width=8 cm,angle=0}}
\caption{The equation of state for deformed radiation for model
1 and 2 (solid lines). The dashed lines represent fits of the 
form $w={1\over 3} +\log(1+D\log(\rho\lambda^4))$ with 
$D\approx 4$ and $D\approx 250$ respectively, and the dotted
line shows the slope 1.
}
\label{w}
\end{figure}

To find the modified equations of state we therefore have
to compute the integral: 
\be\label{eqp}
p(T)={1\over 3}\int  {\rho(E,T) dE\over 1-{f'E\over f}}
\ee
Combined with $\rho(T)$ this leads to a modified
equation of state $p=w(\rho)\rho$. It is not difficult to guess
its form. At low temperatures $w\approx 1/3$. At high temperatures
we have 
\bea
\rho&= & \sigma^+ {T\over \lambda^3}\\
p& \approx & {\sigma^+T\over 3\lambda^3}  {\left(A
 + \log (T\lambda)\right)}
\eea
where $A$  is order 1. So, for $T\lambda\gg 1$, we have
\be\label{wapprox}
w(\rho)\approx A+  \log (T\lambda)\approx  B+ \log (\rho\lambda^4)
\ee
that is $w$ grows logarithmically with $\rho$ (and therefore with
$T$). This approximate argument is confirmed by a numerical integration
as shown in Figure~\ref{w}. 

Not surprisingly one may violate all of Hawking's energy conditions
apart from the weak one (the energy density is still always positive).
Indeed a sign of non-commutativity appears to be that it generates
equations of state with $|w|>1$. The fact that the usually sensible 
energy conditions are badly violated should not deter us. 

\begin{figure}
\centerline{\psfig{file=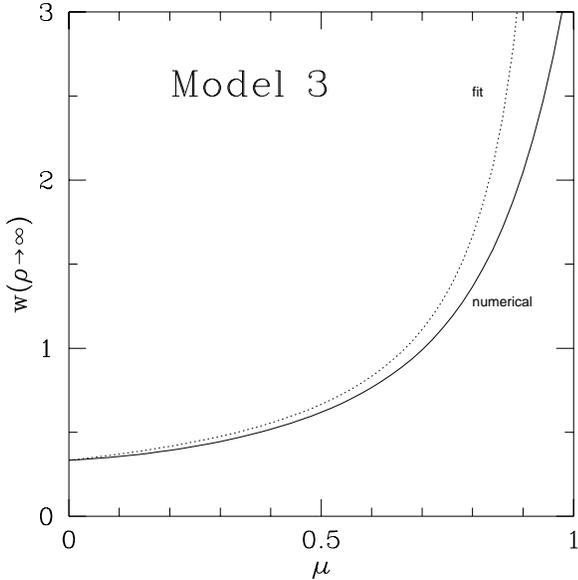,width=8 cm,angle=0}}
\caption{The equation of state for $T\gg \lambda^{-1}$ for model
3 with $\gamma<1$. The dashed line represents the analytical fit
described in the text; the solid line the result of a numerical
integration.
}
\label{w3}
\end{figure}

Model 3 leads to a simpler equation of state. At low energies 
$w=1/3$; at high energies the equation of state becomes again
a constant. Indeed for this model $(f'E)/f=\gamma$ at $E\gg\lambda^{-1}$
so that the
denominator of Eqn.~\ref{eqp} becomes approximately a constant
at high temperatures, over the peak of $\rho(E)$. Hence we may 
expect: 
\be
w_\infty=w(\rho\rightarrow\infty)\approx{1\over 3(1-\gamma)}
\ee
A numerical integration reveals that this is only an approximate
formula, albeit with the correct qualitative behaviour: in Fig.~\ref{w3}
we show the asymptotic equation of state as predicted by our
analytical fit and by a numerical integration. 
In \cite{ncinfl} we shall explore the dramatically
different range $\gamma>1$, to find that it is possible to generate an
inflationary equation of state. 

Notice that the results in this section lead to a closed form expression
for the entropy:
\be
s={p+\rho\over T}={\rho\over T}(1+w(\rho))
\ee
with limiting cases (\ref{lowts}) and (\ref{hights}).

\subsection{The phonon analogy}\label{anal}
We conclude this section by noting that 
to some extent the thermodynamics of deformed radiation is very similar
to that of phonons in crystals. Thermal phonons satisfy Bose-Einstein
statistics; however, unlike photons, they are subjected to very complicated
dispersion relations. Hence the density of states for phonons $\Omega(E)$
(cf. Eqn.~(\ref{densstat})) at high energies is very different from that
of photons (see for instance Fig.8-4, pp 217 of \cite{cond}),  
giving rise to a multitude of thermodynamical novelties similar 
to the ones we have just derived for deformed radiation. 
At low energies the phonons cannot see the discrete structure of the 
crystal and behave like ordinary photons. At high energies, on the contrary,
they become highly sensitive to crystal properties and exhibit exotic
behaviour. The borderline between these 
two regimes, for thermal phonons, is determined by the Debye temperature,
which is therefore
a solid state physics counterpart to  the Planck temperature. 

The analogy we have spelled out is far from surprising. After all
space-time non-commutativity is nothing but a method for quantising 
space-time. In some
sense this means introducing a discrete space-time structure not dissimilar
to that of a crystal (another interpretation is space-time uncertainty
\cite{ram}). As photon frequencies get higher and higher
(or their temperature approaches the Planck temperature) their
dispersion relations, and all derived thermodynamical properties,
start being sensitive to the discreteness of the space-time
structure supporting them.


\section{The trans-Planckian cosmological evolution}

We now proceed to integrate the Friedmann equations with the
$w=w(\rho)$ peculiar to non-commutative radiation. We first
consider the case of zero spatial curvature $K=0$, leaving for
Section~\ref{flat} the cases $K=\pm 1$. The equations 
may be written\footnote{In what follows we shall set $8\pi G=1$}:
\bea
{\left(\dot a \over a\right)} ^2 &=& {1\over 3} \rho 
\label{fried1}\\
{\ddot a \over a}&=& - {1\over 6} \rho(1+3 w(\rho))\label{fried2}
\eea 
where $a$ is the scale factor
and dots represent derivatives with respect to proper time. Instead of 
the second equation one may use the conservation equation
\be\label{cons}
\dot \rho + 3{\dot a \over a}
(1+w(\rho))\rho=0
\ee
From the equations above it is also possible to find 
an equation for the speed of sound $c_s$, defined by:
\be
c_s^2={\delta p\over \delta \rho}={\dot p\over \dot\rho}
\ee
Computing $\dot w$ and using equation (\ref{cons}) leads to
\cite{kodama}:
\be
\dot w= -3{\dot a\over a}(1+w)(c_s^2-w)
\ee
and so
\be
c_s^2=w+\rho{dw\over d\rho}
\ee
an equation which shall be of great relevance in a future publication,
in which we discuss density fluctuations in these scenarios \cite{fluct}.

\begin{figure}
\centerline{\psfig{file=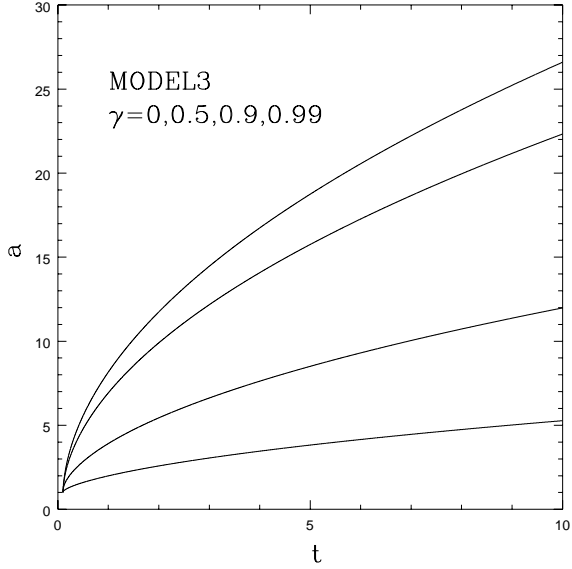,width=8 cm,angle=0}}
\caption{The cosmic evolution (with $t$ and $\rho$ is Planck units)
for Model 3 with various values of $\gamma$. Model 1 can be inferred
from $\gamma\rightarrow 1$. The expansion is always decelerated,
even though the universe expands {\it slower} inside the Planck epoch. }
\label{a2}
\end{figure}
For model 3 it is easy to obtain an analytical solution 
in the asymptotic trans-Planckian regime:
\be\label{aofteq}
a\propto t^{2\over 3(1+w_\infty)}
\ee
A numerical integration (with a $w(\rho)$ also
obtained numerically) is presented in Fig.~\ref{a2}.
We see that after Planck's time the universe expands
as a normal radiation dominated universe, but inside
the Planck epoch it expands slower, the more so the
higher the value of $\gamma$. This is because the decceleration
of the universe is higher:
\be
q_\infty={\ddot a a\over \dot a^2}=
{1\over 2}{2-\gamma\over 1-\gamma}
\ee
Nowhere, in the transition between one regime and the other,
do we have accelerated expansion. This is to be contrasted
with the case $\gamma>1$, explored elsewhere \cite{ncinfl}.

Models 1 and 2 asymptotic trans-Planckian behaviour  may
be inferred from (\ref{wapprox}) and 
(\ref{cons}).  For $\rho\lambda^4\gg 1$:
\be\label{rhoas}
\rho= {\exp{\left(C\over a^3\right)}\over \lambda^{4}}
\ee
where $C$ is a constant. Integrating (\ref{fried1}) then leads to
\be\label{aas}
Ei{\left(C\over 2 a^3\right)}\propto t
\ee
The qualitative behaviour is similar to the one described. A numerical
intergation leads to a result equivalent to model 3 with
$\gamma\rightarrow 1$.

\section{A solution to the horizon problem}\label{horizon}
Naively it may seem self-evident that the horizon problem has been solved
in this model.
As the Universe gets hotter, the photons' average frequency and energy get 
higher, just like in the standard Big Bang model. However in our theory,
unlike in the standard model,
this  results in a larger ``ambient'' speed of light. Hence our model
realises the 
usual VSL solution to the horizon problem \cite{am}, 
not by a direct time-dependence of $c$, but indirectly, via the chain
linking time, temperature, average photon frequency, and $c$.

However, closer scrutiny shows that it is not that simple; in 
fact not all deformations $f$ lead to a solution to the horizon problem. 
On the one hand our analysis confirms that as $t$ decreases below
Planck's time, the temperature keeps increasing. This is obvious
for model 3 but not for model 1 or 2. However we see that 
$a$ keeps  decreasing (cf. Fig~\ref{a2} and Eqn~(\ref{aas})),  
and from Eqn.~(\ref{rhoas}) and (\ref{stef}):
\be\label{temp}
T= {1\over \lambda\sigma^+}\exp{\left(C\over a^3\right)}
\ee
implying that the temperature $T$ does diverge in the early Universe
for all models considered.

However,  as $T$ grows above $\lambda^{-1}$, we find that for model
1 and 2 (but not for model 3, with $\gamma<1$) the color temperature $T_c$
(defining the peak of $\rho(E)$) saturates at $T_c=\lambda^{-1}$.
Hence a hotter plasma does not contain more energetic photons at the
peak of the distribution; it only contains more photons 
(proportionally to $T^2$) at a peak located at the same energy.
Hence the most abundant photons in pre-Planck times will experience
a roughly constant speed of light. Given that we do not have accelerated
expansion, at first it looks as if these two VSL scenarios do not
actually solve the horizon problem. 

Fortunately a further subtlety comes into play. Recall that for
$T\gg \lambda^{-1}$ we have:
\be
\rho(E)={1\over \pi^2\hbar ^3 c^3}
{E^2 T\over f^3}{\left(
1-{f'E\over f}\right)}
\ee
up to $E\approx T$. For $E>T$ the distribution is then dominated by
the exponential cut off $e^{-\beta E}$:
\be
\rho(E)={1\over \pi^2\hbar ^3 c^3}
{E^3e^{-\beta E} \over f^3}{\left(
1-{f'E\over f}\right)}
\ee
For model 2, $\rho(E)$ falls off exponentially away from the peak
at $T_c\approx \lambda^{-1}$,
but for model 1 the fall-off is merely power-law. More concretely
for model 1 there is a non-negligible density of photons with $E\approx T$, 
the fastest photons in the Universe, given by 
\be
\rho_{max}\approx \rho(T)T\approx {T^4\over 1+ (\lambda T)^4}
\approx \lambda^{-4}
\ee
The conclusion is that even though the color temperature saturates
for $T\gg \lambda^{-1}$, there is still a constant density (of
the order of the Planck density)  of photons with energy 
of order $E\approx T$. These are the fastest photons in the Universe,
experiencing a speed:
\be
{\hat c}=c(1+\lambda T)=c{\left(1+{\lambda^4\rho\over
\sigma^+}\right)}\propto \exp{\left(C\over a^3\right)}
\ee
or a similar expression for ${\tilde c}$. 
Fast photons at the tail at $E=T$ ensure causal
contact at a given time. It is therefore
essential that interactions exist between photons at $E=T$ and  
at $E=\lambda^{-1}$. 

A similar calculation for model 2, on the other hand, leads to:
\be
\rho_{max}\approx \rho(T)T\approx T e^{-2\lambda T}
\ee
implying that the relevant photons become exponentially suppressed.
Model 3 on the other hand does not suffer from any of these problems.
Nonetheless the conclusion remains that
 not all deformations $f$ lead to a solution to the horizon problem
in this VSL scenario; but some do. Let us concentrate on model 1 and 3.

\begin{figure}
\centerline{\psfig{file=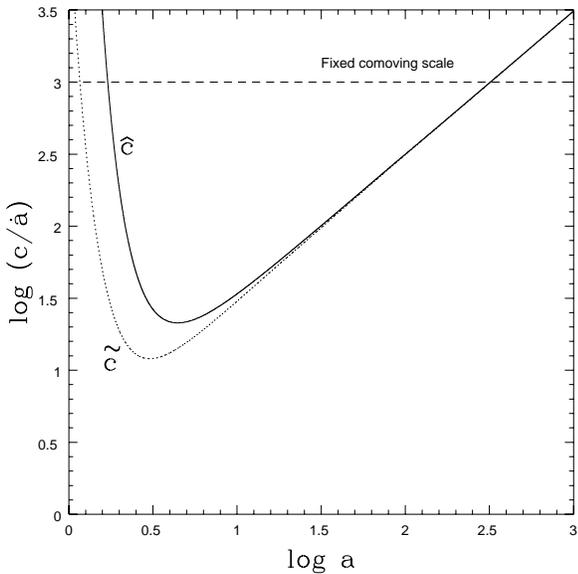,width=8 cm,angle=0}}
\caption{A log-log plot of the comoving horizon distance 
$r_h=c/\dot a$ as a function of the expansion factor $a$.
At late times, as $c$ stabilizes and $a$ acquires the standard
$a\propto t^{1/2}$ behaviour, we recover $r_h\propto a$. 
Before Planck's time, however, the decreasing ${\hat c}$ (or
${\tilde c}$) forces $r_h$ to decrease, thereby solving the horizon
problem.}
\label{hor}
\end{figure}
The causal range at time $t$ in model 1 is defined by the distance 
travelled by photons with $E\approx T$ during one
expansion time (the period over which expansion can be neglected).
The comoving causal range is therefore:
\be
r_h={{\hat c}H^{-1}\over a}={{\hat c}\over \dot a}
\ee
At late times (when $T\lambda\ll 1$),  $c$ stabilizes, and $a\propto
t^{1/2}$, leading to the growth of the comoving horizon $r_h\propto 
a$. Hence the comoving region we can now see was once split into
many disconnected regions. For $T\lambda\gg 1$ we find instead:
\be
r_h={{\hat c}\over \dot a} \propto {\lambda^4\rho\over a{\sqrt \rho}}
\propto {\exp{\left(C\over 2a^3\right)}
\over a}
\ee
a decreasing function of $a$: hence providing a solution to the horizon
problem. 

In Fig.~\ref{hor} we plotted $r_h$ as computed from 
a numerical integration for model 1. We have plotted the
horizon as defined by ${\hat c}$ and by ${\tilde c}$. In both
cases we see the distinctive late time growth $r_h\propto 
a$, but the early exponential decrease in $r_h$. We have plotted
how a given comoving scale is initially in causal contact, then
falls out of causal contact as the speed of light slows down,
to reenter the Hubble radius later on.

For model 3 the solution to the horizon problem is more
straightforward. Since the color temperature never saturates,
as radiation gets hotter so does the ambient speed of light;
we find that
\be
{\hat c}\propto {\tilde c}\propto (\lambda T)^\gamma
\ee
without any extra assumptions. Hence, using $\rho\propto T^{4-3\gamma}$ 
and (\ref{aofteq}) we find
\be
r_h=T^{3\gamma -2\over 2}
\ee
so that for $\gamma>2/3$ we solve the horizon problem.

\section{A solution to the flatness problem}\label{flat}
The discussion in Section~\ref{nonc}
leading to the establishment of a non-commutative
Friedman metric, and equations, breaks down if we introduce spatial
curvature $K=\pm 1$. In that case our space-time is conformally
related not to non-commutative Minkowski space-time, but to a
non-commutative (pseudo-)sphere, namely the fuzzy 
three-sphere $\tilde{S^{3}}$.  For a
derivation of $ \tilde{S^{3}}$ see \cite{san}.

We know that such a fuzzy sphere has the usual properties of 
ordinary spheres as long as the curvature radius is much larger 
than the Planck length. However as one tries to curve them 
beyond  Planck curvature, their curvature
radius saturates. Hence we avoid a singularity. This picture,
however, will not provide a solution to the flatness problem. 
Effectively we may model this effect by replacing the curvature $K/a^2$
by $Kg^2/a^2$, where $g(a)$ describes a deformation of the curvature
of the sphere as it approaches Planck scale. A suitable
function is:
\be\label{defK}
g(a)={1\over 1+{\lambda\over a}}
\ee
and we may check that:
\be
{Kg^2\over a^2}\rightarrow {1\over \lambda^2}
\ee
for $a\ll \lambda$. It is clear that this model would convert
the curvature term into a cosmological constant term, leading to
a pre-Planck deSitter phase which then decays into a curvature
dominated phase. 

We therefore adopt a different approach: we model the fuzzy sphere's
unusual curvature by means of a direct coupling between $K$ and the 
matter density $\rho$. The idea is that the curvature of spatial
surfaces depends upon the energy scale of the matter probing it. This
creates an intertwining between matter and geometry, which is to be
expected above Planck scale. Indeed one does not expect the usual division
between matter and curvature to survive a quantum gravity epoch
(see discussion in the second to last 
paragraph of Section~\ref{nonc}). 
Hence we replace the curvature term $K/a^2$ 
by $Kg^2/a^2$, where now $g=g(\rho)$ describes a deformation of the 
curvature of the sphere as the matter filling up the Universe approaches 
Planck scale. We choose:
\be\label{defK1}
g(\rho)={\left(1+\rho\lambda^4\right)}^\alpha
\ee

As it turns out this approach provides a direct realization of the 
solution to the flatness problem described in \cite{am} (just replace
the $c$ in Friedmann equations by $c_K=cg(\rho)$). 
The Friedmann equations are now:
\bea
{\left(\dot a \over a\right)} ^2 &=& {1\over 3} \rho 
-{Kg^2c^2\over a^2}
\label{fried1b}\\
{\ddot a \over a}&=& - {1\over 6} \rho(1+3 w(\rho))\label{fried2b}
\eea 
which lead to the integrability condition:
\be\label{cons1}
\dot \rho + 3{\dot a \over a}
(1+w(\rho))\rho={6 Kc^2 g^2\over a^2}{g'\rho\over g}{\dot \rho\over \rho}
\ee
(where ' means a derivative with respect to $\rho$) or instead:
\be\label{cons1b}
\dot \rho {\left( 1- {6 Kc^2 g g'\over a^2}\right)}
=-3{\dot a \over a}
(1+w(\rho))\rho
\ee
We find therefore a coupling between the curvature $K$ and the matter
density. In order for flatness to be stable,  $g'$
should be positive: then supercritical models ($K=1$) have matter
removed from them, sub-critical models ($K=-1$) have matter put into them,
thereby creating a flatness attractor. 

\begin{figure}
\centerline{\psfig{file=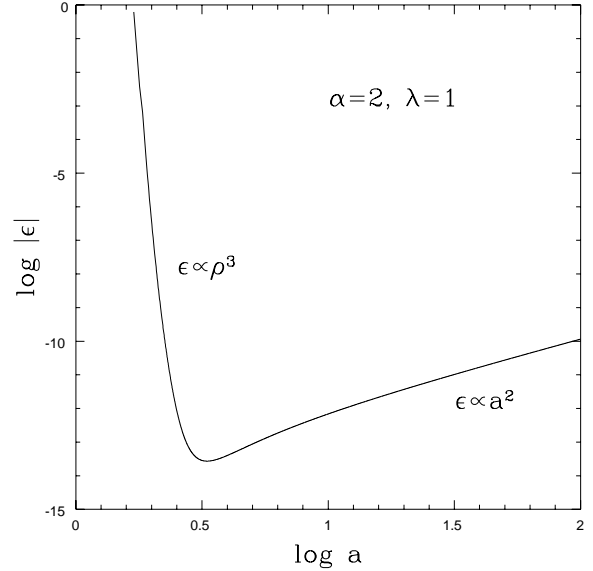,width=8 cm,angle=0}}
\caption{A log-log plot of $|\epsilon |$ versus $a$. We see
that at late times, when the effects of non-commutativity
have switched off $\epsilon\propto a^2$ (the flatness problem),
but early on $\epsilon$ decays like a power of $\rho$.}
\label{flatfig}
\end{figure}

Let $\rho_c$ be the critical density of the Universe:
\begin{equation}
\rho_c={3\over8\pi G}{\left(\dot a\over a\right)}^2
\end{equation}
that is, the mass density corresponding to $K=0$
for a given value of $\dot a/a$. Let us define
$\epsilon=\Omega-1$ with $\Omega=\rho/\rho_c$. 
A numerical integration of equations (\ref{fried1b})
and (\ref{cons1}), expressing the result in terms
of $\epsilon(a)$, is plotted in Fig.~\ref{flatfig}.
We see that at late times, when the effects of non-commutativity
have switched off, $\epsilon\propto a^2$ (the flatness problem),
but early on $\epsilon$ decays very fast, producing a very flat
Universe at the end of the Planck epoch.

This result can be analytically understood by combining
Eqns. (\ref{fried1b}) and (\ref{cons1}) as in \cite{am} 
into a single equation for $\epsilon$:
\begin{equation}\label{epsiloneq}
\dot\epsilon=(1+\epsilon)\epsilon {\dot a\over a} 
{\left(1+3w\right)}+2{\dot g\over g}\epsilon
\end{equation}
Hence, for $\epsilon \ll 1$, and for $\rho\lambda^2\gg 1$, we have
\be
\epsilon\propto a^{1+3w}g^2\approx \rho^{2\alpha -1}
\ee
showing that any exponent $\alpha>1/2$ leads to a solution to the flatness
problem.

For illustration purposes in Fig.~\ref{flatfig} we started the integration at 
$T\approx 10^5 T_P$, and with $\epsilon\approx 1$. In our scenario, however,
there is no starting time, subjected to ``natural'' initial conditions
(we recall that in the Big Bang model one likes to
impose $\epsilon\approx 1$ at $t=t_P$). Our model plunges as deep as
we like inside the Planck epoch, and as we have shown, we found that
during this period flatness then becomes an attractor. Should this regime
be valid all the way up to $T=\infty$ we may conclude that when Planck
time is finally reached $\epsilon$ equals precisely zero. 

\section{The entropy of the Universe}\label{entsection}
The flatness problem is often rephrased as an entropy problem,
both in its inflationary and VSL formulations. As explained
in \cite{infl1,turner,am} adiabatic solutions to the flatness problem lead to
a seemingly very flat Universe, but containing only one particle 
within each curvature radius' volume.
This problem afflicts Brans-Dicke
based solutions to the flatness problem \cite{jana,turner}

All our work assumes thermal equilibrium, so it would seem that we
have an adiabatic scenario - that is, no entropy is produced. 
However this is not the case: we  have introduced source terms
in the energy conservation equation (cf. Eqn.~(\ref{cons1b})), and
this necessarily
modifies the first law of thermodynamics, a fact which,
as we now show, leads to entropy generation even without leaving 
thermal equilibrium.

Taking the equilibrium expression
\be
Td(sa^3)=d(\rho a^3) +pda^3= a^3 d\rho +(\rho +p)da^3
\ee
one may read off from (\ref{cons1}) that
\be
T(sa^3)\dot{}= a^3 {6 Kc^2 g^2\over a^2}{\dot g \over g}
\ee
that is:
\be\label{entropy}
\dot s +3{\dot a \over a}s= {6Kc^2 g\over Ta^2}\dot g
\ee
We see that if $K=0$ entropy is conserved, a fact which may be
checked directly from Eqns.~(\ref{hights}) and (\ref{temp}).
However, if $K=-1$ and $\alpha>0$, then entropy is produced
even though we have only used equilibrium thermodynamics. The
case $K=1$ and $\alpha>0$ seems to lead to entropy reduction,
and thus it may be argued that it is inconsistent with the second
law of thermodynamics (see \cite{chimento} for such an argument in
a different context).

We may now define the entropy (or number of particles) inside
a volume with the curvature radius as
\be
\sigma_K= s{\left(a\over g\right)}^3
\ee
and from (\ref{entropy}) we have:
\be
{\dot \sigma_K\over\sigma_K}=
{\dot g\over g} {\left( {6a\over gT\sigma_K }-3\right)}
\ee
As long as $K=-1$ and $\alpha>0$ we see that $\sigma_K$ increases
unboundedly, under general conditions, before  Planck time. After
Planck time it stabilises to a constant. 
As in the discussion at the end of Section~\ref{flat} we therefore conclude
that the natural state at the end of the Planck epoch is $\sigma_K=\infty$,
explaining the current bound that $\sigma_K$ must be bigger that
$10^{98}$.

The entropy problem is sometimes referred to as the need to explain
why the Universe is so big, or why it contains so much entropy. Usually
one needs to invoke an entropy production mechanism, such as reheating
at the end of inflation, in order to solve this problem.
In our scenario the explanation is related
to the fact that in the non-commutative phase of the Universe we
break Poincar\'e invariance. We know that energy conservation follows 
from time-translation invariance - a property which is now deformed.
Hence we expect violations of energy conservation, and these lead to entropy
production even if we remain in thermal equilibrium.

\section{Conclusions}

There are many avenues which lead to the possibility that space-time at
the Planck scale is non-commutative.  In the context of the Early
Universe, VSL is able to solve the outstanding problems of SBB.  In this
paper we demonstrate that non-commutativity, in the context of the early
universe, is equivalent
to VSL.  We provide a concrete cosmological model of a non-commutative
early universe scenario which exhibits VSL phenomena.  In this model
the speed of light does not depend explicitly on time; rather it
depends on the photon's  frequency, and so as the Universe gets hotter
the ``ambient'' speed of light increases. We showed 
that in this model the horizon and flatness problems are resolved.  

Perhaps the most important result of all, however, is the discovery
that our solution to the horizon and flatness problems is related to
the emergence of Lorentz invariance in the Universe. We found that 
photons in non-commutative space-times are very similar to phonons
in crystals, with Planck's temperature playing the role of the Debye 
temperature (see Section~\ref{anal}). 
At low energies the photons perceive a Lorentz invariant
continuum, and have a constant speed. This corresponds to the standard
Big Bang phase of the Universe. At high frequencies, 
however, the photon's dispersion relation reflects the 
structure and properties of the crystal. The crystal breaks Lorentz
invariance, and like in the case of phonons, the photon's speed becomes
frequency dependent. It is in this phase that the horizon and flatness
problems are solved. The fact that the Universe is pushed towards a colder
phase, for which non-commutativity and violations of Lorentz invariance
become negligible, can be seen as an explanation for the emergence
of the continuum and symmetries we perceive today.

In a future publication \cite{fluct} we shall go further and prove that 
thermal fluctuations in deformed radiation may translate
into a Harrison-Zeldovich spectrum of initial conditions for structure 
formation. Crucially this rules out all but a well-defined class of quantum
deformations. This important result provides a thermal alternative
to the usual quantum origin of the structure of the Universe
(which may or may not have its problems \cite{rob,star}). It also, 
for the first time, allows VSL to become a proper model of 
structure formation. Another avenue to be studied further concerns 
the cosmological constant
problem. Moffat has recently \cite{moflambda}
shown how a resolution may emerge from
non-local field theories. Non-commutativity imposes just such
a non-locality.

In a somewhat different direction in \cite{ncinfl} we describe
our findings for model 3, with $\gamma>1$, a range left untouched
in this paper. We find that ordinary thermal radiation
subject to this type of non-commutativity may drive inflation.
We thus realize the inflationary scenario without the aid of an inflaton field.
As the radiation cools down below Planck's temperature, inflation 
gracefully exits into a standard Big Bang universe, dispensing with 
a period of reheating. Curiously in this regime there is no VSL,
as the color temperature is found to saturate.

Recently, progress in String/M theory and Spin Networks have realized a
non-commutative phase.  We have conjectured a non-commutative version of
the Friedmann-Walker space-time in our realization of the equivalence of
VSL and $\kappa$ deformed non-commutativity in order study the dynamics
the early universe.  We expect that this type of non-commutativity to
come from a non-constant B field in a curved background in
String-theory.  We appreciate that this is beyond the state of the art
in string-theory.  However, it would be important to find a more
concrete realization of NCFRW from string/M theory. 


\section*{Acknowledgements} We would like to thank K. Baskerville,
R. Brandenberger, D. Eagle, H. Jones, and J. Kowalski-Glikman for help with 
this project.

\end{document}